\documentclass[%
 reprint,
 amsmath,amssymb,
 aps,
]{revtex4-1}

\usepackage{graphicx}% Include figure files
\usepackage{dcolumn}% Align table columns on decimal point
\usepackage{bm}% bold math
\begin{document}

\setlength\parskip{\baselineskip}

\title{CPT tests with the antihydrogen molecular ion}
\author{Edmund G. Myers}
\received{2 January 2018}
\affiliation{Department of Physics, Florida State University, Tallahassee, Florida 32306-4350, USA}

\begin{abstract}
   High precision radio-frequency, microwave and infrared spectroscopic measurements of the 
   antihydrogen molecular ion $\bar{H}_{2}^{-}$ ($\bar{p}\bar{p}e^{+}$) compared with its normal matter counterpart provide 
   direct tests of the CPT theorem. 
   %The fractional precision that can be achieved with such measurements
   %can exceed that from comparing antihydrogen with hydrogen, and the
   %$m(e^{+})/m(\bar{p}) - m(e^{-})/m(p)$ can be higher by several 
   %orders of magnitude.
   The sensitivity to a difference between the positron/antiproton and electron/proton mass 
   ratios, and to a difference between the positron-antiproton and electron-proton hyperfine 
   interactions, can exceed that obtained by comparing antihydrogen with 
   hydrogen by several orders of magnitude.
   Practical schemes are outlined for 
   measurements on a single $\bar{H}_{2}^{-}$ ion in a cryogenic Penning trap, that use non-destructive state 
   identification by measuring the cyclotron frequency and bound-positron spin-flip frequency; and also 
   for creating an $\bar{H}_{2}^{-}$ ion and initializing its quantum state.
\end{abstract}

\maketitle

Violation of the CPT theorem, which postulates invariance under the combined 
transformations of charge conjugation, parity and time-reversal would have profound consequences for 
all quantum field theories and fundamental physics \cite{Lueders57,Bluhm99,Yamazaki13,Safranova17}.  A consequence
of CPT is that the properties of fundamental 
particles and their antimatter conjugates should be identical except for the reversal of certain quantum 
numbers. This has led to much effort to compare 
precisely the masses and magnetic moments of the electron and positron \cite{Fee93,VanDyck87,Hanneke08}, 
and of the proton and antiproton \cite{Gabrielse99,Ulmer15,DiSciacca13,Smorra17,Hori16}, and even greater 
efforts to compare the properties of hydrogen and antihydrogen 
\cite{Gabrielse87,Holzscheiter04,Bertsche14,Fitzakerley16,Ahmadi181s2s,Ahmadi17HFS, Kuroda14}.

In the case of antihydrogen ($\bar{H}$), the aim is to measure the $1s_{1/2}$ to 
$2s_{1/2}$ transition by two-photon (2E1) laser spectroscopy, and the $1s_{1/2}$ groundstate
hyperfine splitting (HFS) by microwave spectroscopy, as well as to search for gravitational anomalies 
\cite{Kellerbauer08,Perez15}. 
For the $1s-2s$ transition, the $\bar{H}$-$H$ comparison is sensitive to the 
difference $q(e^{+})^{4}m(e^{+}) -  q(e^{-})^{4}m(e^{-})$, where $q(e^{\pm})$, $m(e^{\pm})$ are the 
respective charges and masses of the positron and electron. While there is also sensitivity to 
$m(e^{+})/m(\bar{p}) - m(e^{-})/m(p)$ through the reduced mass correction, this is decreased by a factor
of 1/1836. In the case of the HFS, the comparison is sensitive to a difference in the product of the 
positron(electron) and antiproton(proton) magnetic moments. $\bar{H}$ has the 
attraction of the possibility of very high precision: for $H$, using cryogenically cooled beams, a
fractional uncertainty of $4 \times 10^{-15}$ has been achieved for the $1s-2s$ transition
\cite{Parthey11}, and $2.7 \times 10^{-9}$ for the $1s$ HFS transition \cite{Diermaier17}.
%using a maser, $7 \times 10^{-13}$ \cite{Petit80}.
However, besides the difficulties of making $\bar{H}$, which currently proceeds by combining 
antiprotons from the Antiproton Decelerator (AD) at CERN \cite{Maury97} with positrons in nested Penning 
traps \cite{Gabrielse88}, experiments with $\bar{H}$ suffer from the difficulty that it must be isolated 
from ordinary matter. Hence, spectroscopic experiments on $\bar{H}$ use weak, large ($\gtrsim$100 cm$^{3}$) 
volume, neutral atom traps, such as the Ioffe-Pritchard trap \cite{Gabrielse08,Andresen10, Richerme13}, or tenuous
beams \cite{Kuroda14}, which pose difficulties for high precision. These include very low densities, 
inhomogeneous magnetic fields, Doppler shifts, and short transit times. So, although the first measurements 
of the $1s-2s$ \cite{Ahmadi181s2s} and $1s$ HFS \cite{Ahmadi17HFS} transitions in $\bar{H}$ have already 
been made, and major improvements can be expected from laser cooling 
\cite{Walz03, Donnan13}, the precision achieved in hydrogen will not be reached for some years.

In contrast to the difficulties of confining antihydrogen, antiprotons (and normal matter ions) have long 
been trapped \cite{Gabrielse86,BG86} and are now routinely manipulated within \cite{Gabrielse99,Rainville04,Myers15} and between \cite{Gabrielse01,Smorra17,Sellner17} cryogenic Penning traps for periods of many months, and they can be tightly confined, 
and their motions precisely monitored using image-current techniques \cite{BG86,Myers13}. This encourages 
consideration of testing CPT by performing precise spectroscopy on the antihydrogen molecular ion 
$\bar{H}_{2}^{-}$, the simplest antiprotonic ion with discrete energy levels. Here it is shown, using 
non-destructive single ion detection techniques, that high-precision measurements on $\bar{H}_{2}^{-}$ are 
possible. Specifically, methods are outlined for the measurement of bound-positron spin-flip (Zeeman) frequencies, 
Zeeman-hyperfine frequencies, and vibrational frequencies using Penning traps, 
that could enable tests of CPT using $\bar{H}_{2}^{-}$ that are several orders-of-magnitude 
more sensitive than can be obtained with bare antiprotons or antihydrogen. This sensitivity 
advantage is particularly great for measurements of the hyperfine interaction, due to the long
coherent interrogation times enabled by a Penning trap; and for measurements of vibrational 
transitions, which are inherently $\sim10^{3}$ more sensitive to
$m(e^{+})/m(\bar{p}) - m(e^{-})/m(p)$ than $1s-2s$ spectroscopy in $\bar{H}$ and $H$.

%that may enable higher 
%fractional precision to be obtained with $\bar{H}_{2}^{-}$ than can be obtained with the bare %particles or 
%$\bar{H}$. In particular, the proposed ultra-high precision comparison of vibrational %frequencies in $\bar{H}_{2}^{-}$ and 
%$H_{2}^{+}$ has a sensitivity to $m(e^{+})/m(\bar{p}) - m(e^{-})/m(p)$ several orders of %magnitude higher 
%than can be envisaged from $1s-2s$ spectroscopy in $\bar{H}$ and $H$. 

\textit{Energy levels:} The
$H_{2}^{+}$ ($\bar{H}_{2}^{-}$) ion in its ground electronic state $1s \sigma_{g} (X ^{2}\Sigma_{g}^{+})$ is strongly bound (dissociation energy $D_{0}$ = 2.6507 eV),
with 20 bound vibrational levels (quantum number $v$) and 423 bound rotational levels 
(quantum number $N$) \cite{Carrington89,Moss93,Brown03}.
The
vibrational level spacing is 65.7 THz for ($v,N$) = (0,0) to (1,0). The number of bound rotational levels 
for each vibrational level decreases from 35 for $v = 0$, to 2 for $v = 19$. The rotational levels have para (ortho) exchange symmetry, with total nuclear spin $I = 0$(1), for $N$ even(odd).  
Because para-ortho transitions are strongly forbidden, the $N$ even and $N$ odd ions are effectively 
separate species. For $N$ even, the HFS is due only to the electron-spin molecular-rotation interaction. For
odd $N$, the hyperfine structure is more complicated due to the additional interactions involving nuclear 
spin.  However, in the 1 to 10 tesla magnetic fields of typical Penning traps, the Zeeman structure is in 
the strong-field regime, and the individual substates can be identified by the projections of the electron 
spin, total nuclear spin (if present), and rotational angular momentum, $M_{S}$, $M_{I}$, $M_{N}$. The Zeeman
splitting with respect to $M_{S}$ is dominant.

Since electric dipole transitions are forbidden in a homonuclear diatomic molecule, 
the ro-vibrational
levels mainly decay by electric quadrupole (E2) transitions with selection rule
$\Delta{N} = 0,\pm2$. Excited vibrational levels have mean lifetimes on the order of
a week or longer \cite{Posen83}.  The rotational levels of $v=0$ 
have mean lifetimes of a few days for $N$ around 30, increasing to 3300 years for $N = 2$ \cite{Pilon12} 
($N$ = 0 to 2 spacing 5.22 THz), while the radiative decay of $N$ = 1 is forbidden.  
Hence $H_{2}^{+}$($\bar{H}_{2}^{-}$) has an abundance of transitions with extremely 
narrow radiative widths. This gives it an advantage for ultra-high precision spectroscopy relative to $H$($\bar{H})$, whose 
only metastable levels are the upper hyperfine level of the ground state, and the $2s$ level, which has a 
lifetime of about 1/8 sec. On the other hand, compared to $H$, and to most atomic ions used for optical 
clocks, \cite{Ludlow15,Huntemann16}, the lack of any electric dipole transition poses challenges by 
preventing direct laser cooling and state detection using fluorescence.

\textit{Single $\bar{H}_{2}^{-}$ ion in a Penning trap:} Here we focus on measurements that 
can be carried out in Penning traps that are compatible with current methods for trapping antiprotons 
\cite{Gabrielse01}. In a precision Penning trap 
\cite{BG86,Myers13}, a set of cylindrically symmetric electrodes produces a quadrupolar electrostatic 
potential aligned with a highly uniform magnetic field. A single ion undergoes an axial motion parallel to 
the magnetic field due to the electrostatic potential, and two circular motions perpendicular to the 
magnetic field, the (modified) cyclotron motion and the magnetron motion. Using image-current techniques, the
motions of a single ion can be cooled into thermal equilibrium with a high quality-factor inductor 
maintained at LHe temperature (4.2K) in time scales of ~0.1 to 30 s. In the near future, by directly cooling the inductor with a 
dilution refrigerator, \textit{e.g.}, see \cite{Peil99}, or by using laser-cooled alkaline-earth ions in an 
adjacent trap with shared electrodes \cite{Heizen90, Bohman17}, or laser-cooled anions in the same trap 
\cite{Kellerbauer06, Yzombard15}, the $\bar{H}_{2}^{-}$ ion temperature may be reduced by a further two to three orders of 
magnitude. Because of the difficulty of making $\bar{H}_{2}^{-}$ ions, state detection and preparation 
methods are devised that enable measurements on a single $\bar{H}_{2}^{-}$ to be repeated indefinitely. In 
particular, they do not use annihilation or photo-dissociation. In what follows, $\bar{H}_{2}^{-}$ is referred to with the understanding that 
the same measurements can (and more easily) be made on $H_{2}^{+}$.

$\textit{Bound positron g-factor:}$ The first of the proposed measurements on $\bar{H}_{2}^{-}$, which also
introduces the main state detection technique, is the simultaneous measurement of the bound-positron 
spin-flip frequency and ion cyclotron frequency, whose ratio is proportional to the positron $g$-factor.  
Although the measurement can be performed on an $\bar{H}_{2}^{-}$ in essentially any ($v,N$), it is simplest
to consider the case of $N$ = 0, where there is no hyperfine structure. The Zeeman structure then consists of
two states with $M_{S} = \pm1/2$, separated by the spin-flip frequency $\simeq$ 28.025 GHz/T due to the 
magnetic moment of the positron, which, except for small bound-state corrections, is the same as for the 
free positron. Besides initial state preparation, which is discussed later, the method is identical to that 
already developed with great success for high precision measurements of the bound-electron $g$-factor 
in $^{12}C^{5+}$ \cite{Werth02}, which have now reached a fractional uncertainty less than $3 \times 
10^{-11}$ \cite{Sturm14,Kohler15}.

The apparatus consists of two adjacent Penning traps in a magnetic field of typically 5 tesla, a ``precision
trap'' where the magnetic field is highly uniform, and where the measurement is carried out; and an ``analysis 
trap'', where the magnetic field has an inhomogeneity with a quadratic spatial dependence $B \simeq B_{0} + 
B_{2}z^{2}$ \cite{BG86}. In the analysis trap the positron spin state $M_{S}$ is determined through the 
shift in axial frequency of the ion due the interaction of its magnetic moment with the quadratic field 
gradient. This is known as the continuous Stern-Gerlach technique (CSG) \cite{Werth02,Dehmelt86}. In contrast to 
the extreme difficulty of detecting a spin-flip of a bare anti-proton, which has a 650 times smaller 
magnetic moment  \cite{DiSciacca13,Smorra17,Schneider17}, the spin-flip of a bound positron produces an 
easily detectable change in axial frequency, enabling determination of $M_{S}$ in 1 minute or less, in a 
magnetic field with modest inhomogeneity \cite{Kohler15}. In the precision trap, the cyclotron frequency of 
the ion is measured by monitoring the evolution of the phase of the classical cyclotron motion 
\cite{Cornell89, Sturm11}, while microwaves are applied at the expected spin-flip frequency, to make an 
attempt at inducing a positron spin-flip. 

The measurement protocol consists of making an attempt at a spin-flip in the precision trap, while 
simultaneously measuring the cyclotron frequency, and then transferring the ion to the analysis trap to 
determine if a spin-flip had occurred in the precision trap. The process is repeated to map out spin-flip 
probability as a function of microwave drive frequency. Because the cyclotron frequency of the 
$\bar{H}_{2}^{-}$ in the $(v,N)$ = $(v,0)$ state, $f_{c}$ = 
$(1/2\pi)Bq(\bar{H}_{2}^{-})/M(\bar{H}_{2}^{-}(v,0))$, is measured simultaneously in the magnetic field $B$ 
of the precision trap, the ratio of spin-flip frequency, $f_{s}$, to $f_{c}$ is independent of $B$ and is given by  
\begin{equation}
\frac{f_{s}}{f_{c}} = \left |\frac{\bar{g}_{e}(\bar{H}_{2}^{-}(v,0))}{2}\frac{q(e^{+})}{m(e^{+})}\frac{M(\bar{H}_{2}^{-}(v,0))}{q(\bar{H}_{2}^{-})}\right |
\end{equation}
where $\bar{g}_{e}(\bar{H}_{2}^{-}(v,0)) \simeq 2.002$, is the effective $g$-factor of the bound positron 
(defined so that the magnetic moment is $\bar{g}_{e}(\bar{H}_{2}^{-}(v,0)) \bar{\mu}_{B} M_{S}$, where 
$\bar{\mu}_{B} =  \hbar q(e^{+})/2m(e^{+})$, with $q(e^{+})/m(e^{+})$ the charge-to-mass ratio of the free 
positron), and  $q(\bar{H}_{2}^{-})/M(\bar{H}_{2}^{-}(v,0))$ is the charge-to-mass ratio of the 
$\bar{H}_{2}^{-}$ ion, with allowance for the ro-vibrational energy. If one assumes the equality of 
$q(e^{+})$ and $-q(e^{-})$, of $-q(\bar{H}_{2}^{-})$ and $q(H_{2}^{+})$, and of 
$\bar{g}_{e}(\bar{H}_{2}^{-}(v,0))$ and $-g_{e}(H_{2}^{+}(v,0))$ \cite{VanDyck87},
the $\bar{H}_{2}^{-}$ to $H_{2}^{+}$ comparison is mainly sensitive to $m(e^{+})/m(\bar{p}) - m(e^{-})/m(p)$. Although higher precision can 
be obtained from vibrational spectroscopy on $\bar{H}_{2}^{-}$, see below, an uncertainty of $3 \times 
10^{-11}$ for a comparison of $m(e)/m(p)$ between matter and antimatter would already be competitive with 
the most precise comparisons of the masses of the proton and antiproton \cite{Gabrielse99,Ulmer15} and of 
the electron and positron \cite{Fee93}.

 \textit{Ro-vibrational state and substate identification:} Using a double-resonance technique the CSG technique can 
 be applied more generally to determine the vibrational and rotational state of a simple paramagnetic 
 molecular ion such as $\bar{H}_{2}^{-}$.  In the case of even $N$, the Zeeman-hyperfine energies of 
 $\bar{H}_{2}^{-}$ in the high magnetic field of a Penning trap are given approximately by \cite{Brown03}
\begin{eqnarray}
E(v,N;M_{S},M_{N};B) \simeq E(v,N) - \bar{g}_{e}(v,N) B \bar{\mu}_{B} M_{S}\nonumber\\
+ \bar{g}_{r}(v,N) B \bar{\mu}_{B} M_{N}
+ \bar{\gamma}(v,N) M_{S} M_{N},\qquad							
\end{eqnarray}
where $\bar{g}_{e}(v,N)$ is the bound-positron $g$-factor, $\bar{g}_{r}(v,N)$ is the rotational $g$-factor, 
and $\bar{\gamma}(v,N)$  is the spin-rotation coupling constant, and $B$ is the magnetic field. Hence, 
positron spin-flip transitions, which, in high $B$ have the selection rule $\Delta{M}_{S} = \pm 1$, 
$\Delta{M}_{N} = 0$, have frequencies given by the energy difference 
\begin{equation}
\Delta{E}(v,N,M_{N};B) \simeq \bar{g}_{e}(v,N) B \bar{\mu}_{B}  - \bar{\gamma}(v,N) M_{N}.			
\end{equation}
Now, while the dependence of $\bar{g}_{e}(v,N)$ on $v$ and $N$ is small, $\bar{\gamma}(v,N)$ has an easily 
resolvable dependence. For example, (for $H_2^+$) $\gamma(v,N)$  has the calculated values 42.162, 41.294, 
39.572 and 38.748 MHz, for $(v,N)$ = (0,2), (0,4), (1,2), and (1,4), respectively \cite{Korobov06}.  Hence, 
$v$, $N$, and $M_{N}$ can be identified by determining the microwave frequency at which the positron 
spin-flip occurs, and comparing it with a theoretical value corresponding to the magnetic field,
which can be determined from a measurement of cyclotron frequency.  In most cases, sufficient resolution to 
identify the state could be achieved by inducing the spin-flips in the analysis trap, despite its 
inhomogeneous magnetic field.  If $M_{N}$ = 0, or if it is otherwise necessary to resolve ambiguities, 
additional information is obtained by inducing radio-frequency (RF) rotational hyperfine-Zeeman transitions with selection rules 
$\Delta{M}_{N} = \pm 1$, $\Delta{M}_{S}= 0$. These would be detected by looking for a 
change in frequency of a subsequent positron spin-flip. 
Examples of positron spin-flip transitions and a Zeeman-hyperfine transition are shown 
in Fig. 1.

\begin{figure}
\includegraphics[scale=0.25]{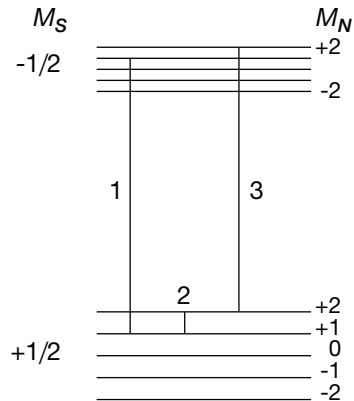}
\caption{Examples of positron spin-flip (Zeeman) transitions, 1, 3, and a rotational  Zeeman-hyperfine transition, 2, in  $\bar{H}_{2}^{-}$, for the case of $N$ = 2 in a magnetic field of 5 T (not to scale). Using the Breit-Rabi formula \cite{Ramsey56} and Zeeman and Hyperfine coefficients from \cite{Karr08Zee,Korobov06}, for $v$ = 0, transitions 1, 2 and 3 have calculated frequencies of 140 082.6, 56.150, 140 040.4 MHz, respectively; and, for $v$ = 1, 140 085.2, 54.499 and 140 045.6 MHz. This illustrates how $M_{N}$ and $v$, (and also, by extension $M_{I}$ and $N$) can be identified by measuring the positron spin-flip frequencies, with additional information given by measuring Zeman-hyperfine transition frequencies.  For ${H}_{2}^{+}$ the level structure is the same, but with the signs of $M_{S}$ and  $M_{N}$ reversed.}
\end{figure}

The case for odd $N$ with $I$ = 1 is more complex, 
with three times as many substates. Nevertheless, the $M_{I}$ state can be identified through the
modification to the positron spin-flip frequency due to the nuclear spin hyperfine interaction, which adds several terms to the effective Hamiltonian, including a term
$b(v,N)M_{S}M_{I}$, where $b(v,N)$ is the
Fermi-contact hyperfine constant.
Again, further identification results by inducing nuclear spin-flip transitions, with selection rules 
$\Delta{M}_{I} = \pm 1$, $\Delta{M}_{S} = 0$, $\Delta{M}_{N} = 0$. To minimize the time 
required for state identification, the microwaves or RF could be applied as pi-pulses, at pre-calculated 
frequencies corresponding to the expected $v$,$N$, $M_{N}$, $M_{I}$ states and the magnetic field.

A complementary method for detecting vibrational transitions, which is especially useful for $N$ = 0 where 
the positron spin-flip frequency is insensitive to $v$, is to make use of the dependence of the cyclotron 
frequency on the vibrational mass-energy of the $\bar{H}_{2}^{-}$. This increases by 1.45, 2.81, and $4.09 \times 
10^{-10}$ for transitions between $v$= 0 and 1, 2, and 3, respectively. Although small, such shifts in cyclotron
frequency are detectable, in a time scale of minutes to an hour, as shifts in a cyclotron frequency 
ratio \cite{Hamzeloui17,Smith18}. This can be implemented by comparison of the cyclotron frequencies of 
the $\bar{H}_{2}^{-}$ and a $D^{-}$ ion in the same precision trap, and then no analysis trap is needed. Alternatively, use can be made of the sensitivity of $f_{s}/f_{c}$ to the $\bar{H}_{2}^{-}$ mass as shown in Eqn. 1., and then another ion is not 
needed. 

\textit{Hyperfine-Zeeman transitions and first-order field independent hyperfine transitions:} Following 
from the above, see Fig.1, precision RF spectroscopy can be carried out on the rotational hyperfine-Zeeman  $\Delta{M}_{N}=
\pm 1$ transitions (for even and odd $N$), and nuclear-spin hyperfine-Zeeman transitions $\Delta{M}_{I}= \pm
1$ (for odd $N$), by making a try at inducing these transitions in the precision trap, and then moving the 
ion to the analysis trap and measuring the positron spin-flip frequency to detect if a change in 
$M_{N}$ or $M_{I}$ occurred. The magnetic field in the precision trap can be calibrated by 
measuring the cyclotron frequency simultaneously, or, by alternating with measurements of positron 
spin-flips. By taking suitable combinations of transitions, the interactions of the rotational and nuclear 
spin magnetic moments, with the external field (Zeeman), and with the positron spin magnetic moment 
(Hyperfine), can be separated \cite{Brown03}. Hence, comparisons of the magnetic moments of the antiproton 
and proton can be made, as has been done for the bare particles \cite{DiSciacca13, Smorra17,Schneider17}, 
but with the advantage of faster detection of the nuclear spin-flips; and also of the HFS, which additionally tests for equality of the antiproton and proton magnetization distributions. Further, for certain $\Delta{M}_{I}= 
\pm 1$ transitions, there are magnetic fields where the hyperfine-Zeeman transition frequencies are 
first-order independent of magnetic field. By adjusting the magnetic field to the appropriate values and 
using Ramsey type excitation schemes \cite{Ramsey56}, this can be exploited to obtain measurements with 
fractional uncertainties less than $10^{-13}$ \cite{Bollinger85}. This fractional precision  is competitive with the most precise 
measurements of hydrogen HFS using masers \cite{Petit80} and is more than three orders of 
magnitide higher than has currently been achieved using a cryogenic beam \cite{Diermaier17}. 

\textit{Ro-vibrational transitions:} In the context of ro-vibrational spectroscopy of $H_{2}^{+}$ in an RF 
trap for fundamental constants and optical clocks, detailed analyses have been made of transition 
probabilities and systematic uncertainties for both 2E1 and E2 transitions,
showing uncertainties can be controlled to the level of $10^{-16}$ or 
below \cite{Hilico01,Karr16,Schiller14PRL,Karr14,Schiller14PRA,Karr08HFS,Karr08Zee,Bakalov14}.
While, for the highest precision, the possibility of trapping and sympathetically 
cooling an $\bar{H}_{2}^{-}$ in an RF trap and
performing quantum logic spectroscopy \cite{Chou10, Chou17,Leibfried17} should also be pursued, in the following it is shown that 
measurements in a Penning trap with uncertainties below $10^{-15}$ are already feasible.

As a specific example, consider the transition (0,2) to (1,2), with  $\Delta{M}_{S} = 0$, $\Delta{M}_{N} = 0$, and $|M_{S} + M_{N}| = 5/2$, \textit{i.e.}, between the “stretched” states, driven as an E2 
transition at 65.4 THz by an ultra-stable laser. Assume that the $\bar{H}_{2}^{-}$  ion is in a 5 T 
precision Penning trap, with axial $(f_{z})$, trap-modified cyclotron $(f_{ct})$ and magnetron frequencies 
$(f_{m})$ near 1 MHz, 35 MHz and 14 kHz, respectively; and that the axial and cyclotron motions are cooled 
using image currents to 20 mK, and the magnetron motion is cooled by magnetron-to-axial coupling \cite{BG86, 
Cornell90} to ~0.3 mK. For transverse laser irradiation, the ion's motion is then in the Lamb-Dicke regime 
with complete suppression of the first-order Doppler shift on the carrier \cite{Wineland98}. The 
second-order Doppler shift leads to a Boltzmann distribution line shape with $e^{-1}$ width of 60 mHz, consistent 
with a fractional uncertainty of $10^{-15}$.  Assuming a laser linewidth $\lesssim$ 0.1 Hz, the transition can 
be induced using a 1 s pi-pulse with intensity of $\sim 6 \mu$W mm$^{-2}$, with a fractional light shift 
of $\ll 10^{-17}$ \cite{Karr14}. The Stark shift, which is mainly due to the cyclotron motion and is proportional to the ion's temperature \cite{Margolis09}, is a factor of $10^{-3}$ smaller than the second-order Doppler shift. The 
Zeeman shift is $1.4 \times 10^{5}$ Hz T$^{-1}$ \cite{Schiller14PRL,Karr08Zee}. But, since a magnetic field stability of better 
than $10^{-9}$ hour$^{-1}$ and calibration to better than $10^{-9}$ can be routinely achieved in 
precision Penning traps, the resulting line broadening and uncertainty are  $<$ 1 mHz. Likewise, the 
quadrupole shift, which is also independent of the ion's temperature, and which can be estimated to be
$\sim1.0$ Hz T$^{-2}$ \cite{Schiller14PRL}, can be calibrated using
knowledge of $f_{ct}$, $f_{z}$ and  $f_{m}$ to better than $10^{-6}$. Hence, besides laser frequency 
stability and metrology, the major limitation to precision is the second-order Doppler shift.

 \textit{$\bar{H}_{2}^{-}$ production and initial state selection:} While it may be possible to synthesize 
 $\bar{H}_{2}^{-}$ in existing or developing antimatter apparatuses using the $\bar{p}  + \bar{H} \rightarrow \bar{H}_{2}^{-} + \gamma$ \cite{Bates51,Zammit17} or 
 $\bar{H}(1s)  + \bar{H}(n\geq 2) \rightarrow \bar{H}_{2}^{-} +  e^{+}$ \cite{Janev, Rawlings94} reactions, $\bar{H}_{2}^{-}$ can be created more robustly through 
 $\bar{H}^{+} + \bar{p} \rightarrow \bar{H}_{2}^{-} + e^{+}$, by merging single cold $\bar{H}^{+}$ 
 ions with a cold $\bar{p}$ plasma. Although $\bar{H}^{+}$ ($\bar{p}e^{+}e^{+}$) has not yet been produced, this is a necessary 
 goal of the ongoing GBAR antihydrogen gravity experiment \cite{Perez15}, in which $\bar{H}^{+}$ will be 
 made by double charge exchange between $\bar{p}$ and positronium, using pulsed $\bar{p}$ and positron beams
 \cite{Walz04}. The GBAR design goal is for one $\bar{H}^{+}$ to be created per AD cycle every 2 minutes 
 \cite{Perez15}. Injected into a $\bar{p}$ plasma with density of $10^{6}$ cm$^{-3}$ at T $\simeq$ 100 K, 
 conditions already achieved \cite{Richerme13,Andresen10}, an $\bar{H}_{2}^{-}$ production 
 rate of $1.4 \times 10^{-3}$ s$^{-1}$ can be estimated \cite{Urbain86,Poulaert78}. However, because of the 180 times 
 larger cross-section of the competing $\bar{H}^{+} + \bar{p} \rightarrow \bar{H} + \bar{H}$, reaction 
 \cite{Stenrup09,Nkambule16}, an $\bar{H}_{2}^{-}$ will be produced on average once 
 per 180 $\bar{H}^{+}$ injections, with mixing times of $\sim$10 s. 

$\bar{H}^{+} + \bar{p} \rightarrow \bar{H}_{2}^{-} + e^{+}$ is exothermic by 1.896 eV. Hence, 
in a cool antiproton plasma the $\bar{H}_{2}^{-}$ will be produced with $v \leq 8$ and $N\leq27$.
By transferring to a higher field (10T) Penning trap, and placing the ion 
in a large radius ($\gtrsim 4$ mm) cyclotron orbit, the vibrational motion can be Stark quenched to $v$ = 0 
through the induced electric dipole moment \cite{Schiller14PRA}, in a time scale $\sim1$ week. Identification of $v$, $N$, $M_{N}$, and $M_{I}$ proceeds by transferring to the analysis
trap and determining the frequency of the positron spin-flip transition, with manipulation of $M_{N}$ and 
$M_{I}$ by hyperfine-Zeeman transitions. Reduction of $N$ (and also $v$, if necessary) can then be 
effected using successive ro-vibrational transitions $(v,N)$ to $(v',N-2)$, again with state 
monitoring via the CSG technique in the analysis trap. The processes of making, state initialization, and measuring on
a single $\bar{H}_{2}^{-}$ may take many weeks. However, such time scales are already common for experiments on single ions in precision Penning traps \cite{Smorra17,Schneider17,Hamzeloui17}. 

\textit{Conclusion:} Precision measurements on $\bar{H}_{2}^{-}$ can provide tests of the CPT theorem that 
are more sensitive than those achievable with antihydrogen or the bare particles, particularly with regards
to the positron(electron)/antiproton(proton) mass ratio, and the positron(electron) antiproton(proton) hyperfine interaction. Practical schemes have been 
outlined for their implementation based on single-ion Penning trap techniques, including the continuous 
Stern-Gerlach effect and measurement of cyclotron frequency for state identification. The $\bar{H}^{+} +
\bar{p} \rightarrow \bar{H}_{2}^{-} + e^{+}$ reaction has been identified as a practical path for $\bar{H}_{2}^{-}$ 
production.
 
The author thanks G. Gabrielse for enabling visits to CERN, and thanks members of the ATRAP collaboration for their 
hospitality. Support by the NSF under PHY-1403725 and PHY-1310079 is acknowledged.

\end{document}